%% file: main.tex
\documentclass[a4paper,11pt]{article}

\usepackage[
left=2.5cm,
right=2.5cm,
top=2.5cm,
bottom=2.5cm
]{geometry}

\usepackage{graphicx}
\usepackage{booktabs}
\usepackage{amsmath,amssymb}
\usepackage{newtxtext,newtxmath}
\usepackage[ruled,vlined,linesnumbered]{algorithm2e}
\usepackage[authoryear,round]{natbib}
\usepackage{hyperref}
\usepackage{xurl}
\usepackage{microtype}
\usepackage{enumitem}
\usepackage{caption}
\usepackage{multirow}
\usepackage{array}
\usepackage{longtable}
\usepackage{listings}
\usepackage{float}

\newcolumntype{L}[1]{>{\raggedright\arraybackslash}p{#1}}
\newcolumntype{C}[1]{>{\centering\arraybackslash}p{#1}}
\lstdefinestyle{promptstyle}{
	basicstyle=\ttfamily\footnotesize,
	breaklines=true,
	breakatwhitespace=false,
	breakautoindent=true,
	breakindent=1.5em,
	columns=fullflexible,
	keepspaces=true,
	resetmargins=true,
	showstringspaces=false,
	frame=tb,
	xleftmargin=1em,
	xrightmargin=1em,
	aboveskip=0.8em,
	belowskip=0.8em,
	captionpos=t
}
\graphicspath{{figures/}}

\hypersetup{
	colorlinks=true,
	linkcolor=black,
	citecolor=black,
	urlcolor=black,
	pdftitle={A Task-Oriented Multi-Agent Framework for Complex Wearable Health Analysis},
	pdfauthor={Kunpeng Yang}
}

\SetKwInput{KwIn}{Input}
\SetKwInput{KwOut}{Output}
\SetKwComment{Comment}{$\triangleright$\ }{}

\title{A Task-Oriented Multi-Agent Framework for Complex Wearable Health Analysis}
\author{%
	Kunpeng Yang\\
	Renmin University of China\\
	\href{mailto:math_co@163.com}{\texttt{math\_co@163.com}}
}
\date{}

\begin{document}
	
	\pagestyle{plain}
	
	\maketitle
	\thispagestyle{plain}
	
	\input{sections/abstract}

	\input{sections/introduction}
	\input{sections/related_work}
	\input{sections/method}
	\input{sections/synthetic_dataset}
	\input{sections/experiments}
	\input{sections/results_discussion}
	\input{sections/conclusion}

	\clearpage
	\begingroup
	\small
	\setlength{\bibsep}{0pt plus 0.25ex}
	\bibliographystyle{plainnat}
	\bibliography{references}
	\endgroup
	
	\clearpage
	\appendix
	\input{sections/appendix}

\end{document}

%% file: sections/abstract.tex

\begin{abstract}
	Wearable health questions often combine data retrieval, longitudinal analysis, and
	health advice over structured records. Prompting a single large
	language model with a complete record and a composite query obscures whether every
	request is executed and which evidence supports the answer. We propose a
	task-oriented multi-agent framework that represents a composite query as distinct
	intents and typed tasks with explicit intra-intent dependencies. Specialized agents execute
	retrieval, analysis, and advice tasks; isolated intent states preserve request
	boundaries and evidence relationships before aggregation.
	We evaluate the framework on a synthetic dataset of $10{,}000$ virtual users with
	one month of longitudinal wearable records, covering structured data retrieval,
	multi-intent recognition, and overall response quality. Across $1{,}500$ retrieval
	questions, the Query Agent achieves $98.3\%$ accuracy, compared with $97.9\%$ for
	the Direct LLM baseline, while reducing average query-stage token consumption from
	$6{,}869$ to $3{,}136$. On $180$ multi-intent questions, the Manager Agent achieves
	$100.0\%$ Multi-Intent Coverage and $94.4\%$ Multiset Jaccard Similarity. Under the
	current synthetic evaluation setting, our method receives higher mean
	Trustworthiness and Transparency scores on both question categories,
	whereas Actionability does not improve consistently. These results provide
	preliminary evidence that explicit task organization can support task-relevant data access
	and data-grounded longitudinal analysis, while leaving health advice
	generation and validation on real wearable data as open challenges.
\end{abstract}

%% file: sections/introduction.tex

\section{Introduction}
\label{sec:introduction}

Wearable devices produce longitudinal records spanning exercise, daily activity,
sleep, and physiological events. These records can support personalized interpretation
of health changes, but their scale and structure make direct inspection impractical.
Users instead need natural-language interfaces that can retrieve measurements, compare
indicators over time, and provide appropriate health-support information. This paper
studies how to answer \emph{composite wearable health queries}: questions that combine
several requests over structured longitudinal records and require each request to remain
traceable to the evidence used in its answer.

Composite queries create three coupled technical challenges. First, one question may
contain several distinct user requests, including repeated instances of the same intent
type and shared temporal constraints that must be preserved during separation.
Second, an analysis or advice request may require evidence produced by multiple
retrieval operations; execution order alone does not specify which predecessor results
should be available to each downstream step. Third, natural-language time expressions
and health indicators must be mapped to exact dates and fields in the underlying
record. Errors in request coverage, temporal resolution, or evidence selection can then
propagate into longitudinal analysis and advice generation.

Large language models (LLMs) have shown considerable potential for medical question
answering and health consultation~(\citealp{singhal2025toward}), but factual errors and
unreliable outputs remain concerns when tasks involve specialized numerical information
and complex health reasoning~(\citealp{wang2024applications};
\citealp{liu2024large}). Statistical tools, information retrieval, code execution, and
agent mechanisms extend what LLM-based health systems can process. Yet directly
providing a complete longitudinal record and a composite question to one model still
leaves request coverage, structured data access, and evidence flow implicit: the same
generation process must identify each request, locate records, organize numerical
results, and compose the final answer.

The resulting gap is primarily one of execution representation rather than the number
of agents. For a composite wearable query, a system needs to preserve distinct user
requests, express the operations required by each request, and identify the predecessor
evidence consumed by downstream analysis or advice. It must also separate deterministic
access to structured records from language-model generation. Existing progress in
tool-augmented and multi-agent health systems motivates these capabilities, while the
organization of repeated intents, intra-query task dependencies, and structured-record
evidence within a single composite query remains less explicitly studied.

We address this gap with the Task-Oriented Multi-Agent Health Assistance Framework.
The framework represents a composite query through an \emph{intent--task} hierarchy: a
Manager Agent identifies distinct user intents and decomposes each into
typed tasks with explicit dependencies. Dependencies are restricted to tasks within the
same intent, so downstream Analyse and Advice Tasks receive only their specified
predecessor results while each intent maintains an isolated local state. Task types are
deterministically mapped to specialized agents. In particular, the Query Agent resolves
temporal constraints, accesses target records through structured interfaces, and
textualizes only task-relevant evidence for subsequent processing. After all intents
finish, the Manager Agent aggregates their terminal results in the order of the
original question.

Our main contributions are as follows:

\begin{enumerate}[label=(\arabic*),leftmargin=2.8em]

	\item We formulate composite wearable health analysis through an intent--task
	hierarchy that preserves distinct requests and represents predecessor-evidence
	relationships as explicit intra-intent task dependencies. The resulting execution
	structure ties specialized-agent invocation to task semantics rather than to an
	open-ended communication protocol.

	\item We design a Query Agent that separates natural-language temporal resolution,
	structured record access, and answer generation. This design supplies downstream
	language models with task-relevant evidence instead of the complete longitudinal
	record and is evaluated against the Direct LLM baseline using the same
	foundation model.

	\item We construct a synthetic evaluation dataset containing $10{,}000$ virtual
	users and one month of longitudinal wearable records, together with evaluations of
	structured data retrieval, multi-intent recognition, and overall response quality.
	We release the implementation, data-generation pipeline, and synthetic dataset to
	support reproducibility and subsequent research.\footnote{
		The source code, data generation pipeline, and synthetic wearable dataset
		are available at
		\url{https://github.com/yangkunpeng-coder/WearableDeviceAgents}.
	}

\end{enumerate}

The synthetic evaluation examines structured retrieval, multi-intent recognition, and
final response quality. Because it does not isolate task decomposition or dependency
construction through component ablations, we treat these mechanisms as system-design
contributions rather than independently verified sources of performance gains.

%% file: sections/related_work.tex

\section{Related Work}
\label{sec:related-work}

\subsection{Large Language Models for Wearable Health Analysis}
\label{subsec:wearable-health-llm}

LLM-based wearable health systems increasingly connect natural-language interaction
with longitudinal personal records. PH-LLM (\citealp{khasentino2025personal}) generates
personalized sleep and exercise insights, and PhysioLLM
(\citealp{fang2024physiollm}) combines wearable data, user context, and statistical
analysis to explain relationships among personal health indicators. HealthGuru
(\citealp{wang2025healthguru}) further applies a multi-agent mechanism to personalized
sleep-health support. Together,
these studies demonstrate the value of grounding health-oriented language interaction
in personal sensor data and analytical tools.

WEQA is particularly related because it uses an LLM controller to construct
query-adaptive execution plans, compose sensor-analysis tools and pretrained models,
and audit answer evidence against external knowledge~(\citealp{zhang2026weqa}). Its
primary concern is heterogeneous tool and model selection across sensing, analysis, and
prediction tasks. Our work addresses a complementary level of organization: retaining
multiple distinct requests within one composite question and associating each
downstream task with explicit predecessor evidence from structured longitudinal
records. Thus, WEQA emphasizes query-adaptive tool composition, whereas our framework
emphasizes multi-intent representation and intra-query evidence flow.

\subsection{Agent Systems for Personal Health}
\label{subsec:health-agent-systems}

Agentic personal-health systems extend language models with executable tools,
specialized roles, and persistent user context. The system of
\citet{merrill2026transforming} combines multi-step reasoning, code execution, and
information retrieval for iterative analysis of personal wearable data, while the
personal health agent of \citet{heydari2025anatomy} delegates data science, health
expertise, and health coaching to specialized sub-agents. These systems establish useful
patterns for tool-augmented analysis and role specialization.

Other systems focus on longitudinal operation beyond a single interaction. HealthClaw
separates shared safety rules and medical knowledge from private longitudinal memory,
then determines whether interaction information should update the user profile,
reusable workflows, or episodic memory~(\citealp{li2026selfevolving}). HiMe instead
emphasizes self-hosting, privacy, real-time processing, and database-centric long-term
user modeling~(\citealp{liu2026hime}). Our framework does not study cross-session memory
evolution or local deployment. It focuses on the within-query execution structure that
records each intent, its tasks, and the predecessor results supplied to downstream
agents. Specialized roles therefore serve an explicit request--task representation
rather than constituting the contribution by themselves.

\subsection{Task Decomposition and Agent Orchestration}
\label{subsec:task-agent-orchestration}

General-purpose agent research provides several foundations for organizing complex
work. ReAct interleaves language reasoning with external actions
~(\citealp{yao2023react}), AutoGen supports interaction among agents, roles, and tools
~(\citealp{wu2024autogen}), and MetaGPT uses role specialization and structured
workflows for collaborative task execution~(\citealp{hong2024metagpt}). DyLAN further
selects agents dynamically and adapts their interaction structure to the task
~(\citealp{liu2024dylan}). Collectively, these approaches study reasoning--action
interaction, communication, routing, and workflow organization in general settings.

Our framework draws on these orchestration principles but does not introduce a general
communication protocol or learned router. It uses deterministic task-type routing and
an execution representation tailored to structured wearable records. The relevant
distinctions are repeated-intent preservation, natural-language temporal mapping, and
explicit selection of predecessor evidence for downstream analysis and advice, rather
than dynamic agent-team construction.

\subsection{Positioning of This Work}
\label{subsec:related-work-positioning}

Across these directions, prior work has established wearable analysis, tool-augmented
health agents, persistent health memory, local deployment, and general multi-agent
orchestration. We focus more narrowly on organizing a single composite
wearable health query. The intent level preserves distinct user requests;
the task level represents the operations and intra-intent evidence dependencies needed
to fulfill each request; and structured interfaces separate record access from language
generation. The primary distinction is therefore an explicit representation that links
requests, tasks, and evidence in multi-intent longitudinal queries, not a claim of more
agents or a new general-purpose agent architecture.

%% file: sections/method.tex

\section{Method}
\label{sec:method}

We propose a task-oriented multi-agent framework that represents a composite wearable
health query as distinct intents and typed tasks with explicit
predecessor dependencies.

\subsection{Problem Formulation}
\label{sec:problem-formulation}

Let $u$ index a user, $q$ denote a natural-language health query, and
$\mathcal{D}_u$ denote the user's complete structured record, comprising a profile and
date-indexed wearable records. The framework produces a response $y$ from $q$ and
$\mathcal{D}_u$. The profile is extracted rather than supplied independently:
\begin{equation}
    \mathcal{B}_u
    =
    \operatorname{ExtractProfile}(\mathcal{D}_u),
    \label{eq:profile-extraction}
\end{equation}
As formalized in Eq.~\eqref{eq:profile-extraction}, $\mathcal{B}_u$ contains demographic,
lifestyle, and health-history information. Intent recognition returns a sequence ordered
by occurrence in $q$:
\begin{equation}
    \mathcal{I}
    =
    \operatorname{RecognizeIntents}(q)
    =
    \left(i_1,i_2,\ldots,i_N\right),
    \label{eq:intent-sequence}
\end{equation}
Each $i_j=(\mathrm{text}_j,\kappa_j)$ contains a self-contained request description and
type $\kappa_j$ from the shared space
\[
    \mathcal{C}
    =
    \left\{
    \mathtt{query},
    \mathtt{analyse},
    \mathtt{advice}
    \right\},
\]
where $N$ is the number of intents in Eq.~\eqref{eq:intent-sequence}. Shared constraints
are retained in every affected
$\mathrm{text}_j$, and repeated types remain distinct sequence elements. We use
\emph{Analyse Agent} and \emph{Analyse Task} for \texttt{analyse} throughout.

Each intent $i_j$ is decomposed into the ordered task sequence
\[
    \mathcal{T}^{(j)}=(t_1^{(j)},\ldots,t_{M_j}^{(j)}),
\]
where $M_j$ is the number of tasks. Each task is
\begin{equation}
    t_m^{(j)}
    =
    \left(
    \mathrm{id}_m^{(j)},
    \mathrm{desc}_m^{(j)},
    c_m^{(j)},
    D_m^{(j)}
    \right),
    \label{eq:task-definition}
\end{equation}
where $\mathrm{id}_m^{(j)}$ and $\mathrm{desc}_m^{(j)}$ denote the within-intent task
identifier and description, respectively, $c_m^{(j)}\in\mathcal{C}$ is the task type,
and $D_m^{(j)}$ is the predecessor-ID set. Thus, Eq.~\eqref{eq:task-definition}
separates task identity, content, operation type, and dependencies. Intent types describe
requested outcomes; task types describe operations. Hence, an
\texttt{analyse} or \texttt{advice} intent may contain preceding retrieval or analysis
tasks, while its terminal task satisfies $c_{M_j}^{(j)}=\kappa_j$.

Task dependencies may reference only preceding tasks within the same intent:
\begin{equation}
    D_m^{(j)}
    \subseteq
    \left\{
    \mathrm{id}_p^{(j)}
    \mid
    1\leq p<m
    \right\}.
    \label{eq:intra-intent-dependency}
\end{equation}
Eq.~\eqref{eq:intra-intent-dependency} permits only completed same-intent
predecessors; tasks requiring no predecessor use $D_m^{(j)}=\emptyset$.

A Query Task accesses $\mathcal{D}_u$ directly and therefore has no predecessor, as
required by Eq.~\eqref{eq:query-empty-dependency}:
\begin{equation}
    c_m^{(j)}=\mathtt{query}
    \quad\Longrightarrow\quad
    D_m^{(j)}=\emptyset.
    \label{eq:query-empty-dependency}
\end{equation}
Analyse and Advice Tasks instead receive the completed same-intent predecessors named
by $D_m^{(j)}$.

Table~\ref{tab:method-notation} summarizes the notation used throughout this section.
\begin{table}[H]
    \centering
    \caption{Summary of the principal notation used in the method.}
    \label{tab:method-notation}
    \footnotesize
    \setlength{\tabcolsep}{5pt}
    \renewcommand{\arraystretch}{1.05}
    \begin{tabular}{L{3.0cm} L{11.2cm}}
        \toprule
        \textbf{Symbol} & \textbf{Definition} \\
        \midrule
        $u$ & User index. \\
        $q$ & Natural-language health query issued by user $u$. \\
        $\mathcal{D}_u,\mathcal{B}_u$ & Complete structured record and extracted profile of user $u$. \\
        $\mathcal{I},i_j$ & Ordered intent sequence and its $j$-th intent. \\
        $\mathcal{T}^{(j)},t_m^{(j)}$ & Ordered task sequence for intent $j$ and its $m$-th task. \\
        $\mathcal{C},c_m^{(j)}$ & Shared type space and the type of task $t_m^{(j)}$. \\
        $D_m^{(j)},\mathcal{P}_m^{(j)}$ & Predecessor identifiers and the corresponding ordered task descriptions and results. \\
        $\mathcal{R}^{(j)},r_m^{(j)},r^{(j)}$ & Intent-local result store, task result, and terminal intent result. \\
        $K_m,L_m$ & Numbers of findings and recommendations produced by a task. \\
        $\mathcal{R}_{\mathcal{I}},\Gamma,y$ & Ordered intent-result sequence, final response renderer, and final response. \\
        \bottomrule
    \end{tabular}
\end{table}

\subsection{Framework Overview}
\label{sec:framework-overview}

\begin{figure}[htbp]
    \centering
    \includegraphics[width=\linewidth]{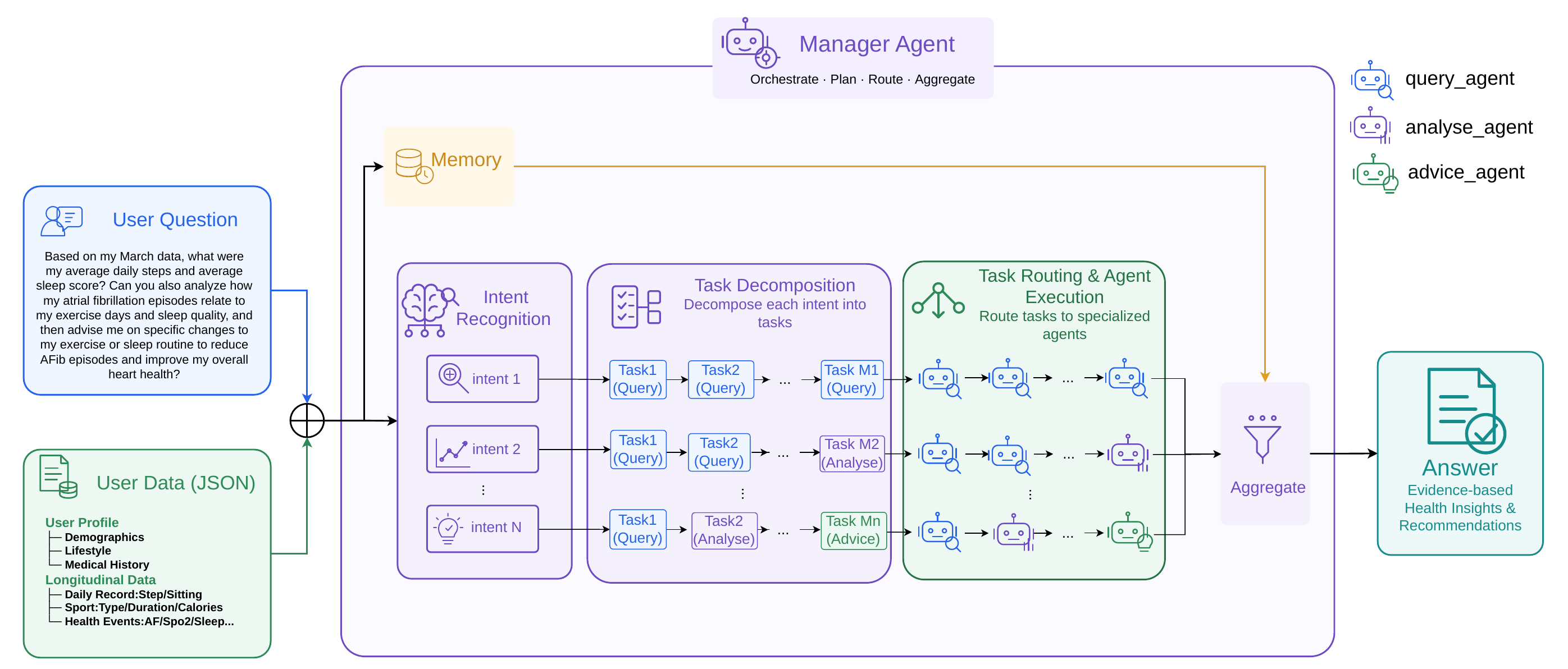}
    \caption{Overview of the task-oriented multi-agent framework. A composite query is decomposed into isolated intents and typed tasks with explicit dependencies. Specialized agents execute the tasks, and the terminal intent results are aggregated into the final response.}
    \label{fig:framework}
\end{figure}

Figure~\ref{fig:framework} summarizes recognition, decomposition,
dependency-conditioned execution, and aggregation.

The Manager Agent recognizes ordered intents, constructs typed tasks and same-intent
dependencies, routes tasks by type, maintains intent-local results, and aggregates
terminal results in query order. Query, Analyse, and Advice Agents execute tasks within
these boundaries, preventing unrelated intermediate results from entering their context.

\subsection{Multi-Intent Recognition and Explicit Task Decomposition}
\label{sec:intent-task-decomposition}

Intent recognition separates distinct user requests into isolated intent-local workflows
while preserving each request's temporal range, indicators, and other constraints. For example, a question
about three-week changes in atrial fibrillation, exercise, and sleep followed by joint
analysis and advice yields three Query Intents, one Analyse Intent, and one Advice
Intent; the shared time range is copied into every affected description.

Decomposition receives only $(\kappa_j,\mathrm{text}_j)$ and generates the current
intent's retrieval, analysis, or advice operations. Even when intents
require the same data, their task structures and dependencies remain separate.

An Advice Intent does not reuse intermediate results from a separate Query Intent. When
record evidence is required, its task sequence contains a new intent-local Query Task,
and the terminal Advice Task explicitly depends on that task's result. This preserves
intent isolation while allowing advice generation to use evidence retrieved for the
advice request itself.

Explicit dependencies represent data relationships among tasks within an intent. For
the current task $t_m^{(j)}$, the framework collects the predecessor results specified
by $D_m^{(j)}$:
\begin{equation}
    \mathcal{P}_m^{(j)}
    =
    \left(
    \left(\mathrm{id}_p^{(j)},\mathrm{desc}_p^{(j)},r_p^{(j)}\right)
    \right)_{\substack{1\leq p<m\\ \mathrm{id}_p^{(j)}\in D_m^{(j)}}},
    \label{eq:dependency-results}
\end{equation}
The ordered triples in Eq.~\eqref{eq:dependency-results} retain source identifiers and
task descriptions with their results; $D_m^{(j)}=\emptyset$ implies
$\mathcal{P}_m^{(j)}=\emptyset$.

In the preceding example, three Query Tasks can feed one Analyse Task whose dependency
set names all three identifiers. Tasks execute in generated order, but $D_m^{(j)}$
determines which completed results the current task receives.

\subsection{Agent Interface Contracts and Routing}
\label{sec:agent-orchestration}

Table~\ref{tab:task-agent-mapping} defines the three execution agents and their input
boundaries. Routing is deterministic by task type; $\mathcal{B}_u$ denotes the profile
extracted from $\mathcal{D}_u$.

\begin{table}[htbp]
    \centering
    \caption{Deterministic mapping from task types to specialized agents and their input boundaries.}
    \label{tab:task-agent-mapping}
    \small
    \setlength{\tabcolsep}{5pt}
    \begin{tabular}{L{2.0cm} L{2.5cm} L{7.6cm}}
        \toprule
        \textbf{Task Type} & \textbf{Agent} & \textbf{Primary Inputs and Responsibilities} \\
        \midrule
        \texttt{query} & Query Agent &
        Complete task, user profile, and complete structured record; resolves temporal constraints from the task description, accesses records on target dates, and produces retrieval results. \\
        \texttt{analyse} & Analyse Agent &
        Complete task, user profile, and explicitly dependent predecessor results; performs longitudinal comparisons and organizes multi-indicator results. \\
        \texttt{advice} & Advice Agent &
        Complete task, user profile, and explicitly dependent retrieval evidence and analysis results; generates health-support advice consistent with the available information. \\
        \bottomrule
    \end{tabular}
\end{table}

The corresponding task-execution contracts are
\begin{equation}
\begin{aligned}
    r_m^{(j)}
    &=
    \operatorname{QueryAgent}
    \left(t_m^{(j)},\mathcal{B}_u,\mathcal{D}_u\right),
    && c_m^{(j)}=\mathtt{query},\\
    r_m^{(j)}
    &=
    \operatorname{AnalyseAgent}
    \left(t_m^{(j)},\mathcal{B}_u,\mathcal{P}_m^{(j)}\right),
    && c_m^{(j)}=\mathtt{analyse},\\
    r_m^{(j)}
    &=
    \operatorname{AdviceAgent}
    \left(t_m^{(j)},\mathcal{B}_u,\mathcal{P}_m^{(j)}\right),
    && c_m^{(j)}=\mathtt{advice}.
\end{aligned}
\label{eq:agent-interface-contracts}
\end{equation}
Eq.~\eqref{eq:agent-interface-contracts} restricts $\mathcal{D}_u$ to the Query
Agent; the other agents receive only selected predecessors. After execution,
\begin{equation}
    y
    =
    \Gamma\left(q,\mathcal{R}_{\mathcal{I}}\right),
    \label{eq:response-aggregation}
\end{equation}
where $\mathcal{R}_{\mathcal{I}}$ is the ordered terminal-result sequence defined in
Section~\ref{sec:advice-aggregation}, and $\Gamma$ is the final LLM-based response
renderer. It organizes terminal results in the context of $q$ without re-accessing
$\mathcal{D}_u$ and is instructed to use only the supplied terminal results without
introducing unsupported record evidence.

\subsection{End-to-End Dependency-Aware Execution}
\label{sec:end-to-end-execution}

Algorithm~\ref{alg:multi-agent-execution} instantiates the preceding representation and
interfaces with an intent-local result store for each intent.
$\mathcal{R}^{(j)}$ denotes this intent-local result store.
$\operatorname{ValidateTaskGraph}$ checks task types, unique identifiers, valid
same-intent predecessor references, empty Query-Task dependencies, and terminal-type
consistency before execution. Thus, only listed same-intent predecessors reach
downstream tasks, and only terminal intent results reach global aggregation.

\begin{algorithm}[tbp]
    \caption{End-to-end dependency-aware execution.}
    \label{alg:multi-agent-execution}
    \small

    \KwIn{Natural-language health query $q$ and complete structured record $\mathcal{D}_u$}
    \KwOut{Final response $y$}

    $\mathcal{B}_u\leftarrow\operatorname{ExtractProfile}(\mathcal{D}_u)$;
    $\mathcal{I}\leftarrow\operatorname{RecognizeIntents}(q)$\;
    Initialize intent-level result sequence $\mathcal{R}_{\mathcal{I}}$\;

    \For{$j=1,\ldots,N$}{
        $\mathcal{T}^{(j)}\leftarrow
        \operatorname{DecomposeTasks}(\kappa_j,\mathrm{text}_j)$\;
        $\operatorname{ValidateTaskGraph}(\mathcal{T}^{(j)})$\;
        Initialize local result store $\mathcal{R}^{(j)}$ for the current intent\;

        \For{$m=1,\ldots,M_j$}{
            \eIf{$c_m^{(j)}=\mathtt{query}$}{
                $r_m^{(j)}\leftarrow
                \operatorname{QueryAgent}
                (t_m^{(j)},\mathcal{B}_u,\mathcal{D}_u)$\;
            }{
                $\mathcal{P}_m^{(j)}\leftarrow
                \operatorname{CollectResults}
                (D_m^{(j)},\mathcal{R}^{(j)})$\;

                \uIf{$c_m^{(j)}=\mathtt{analyse}$}{
                    $r_m^{(j)}\leftarrow
                    \operatorname{AnalyseAgent}
                    (t_m^{(j)},\mathcal{B}_u,\mathcal{P}_m^{(j)})$\;
                }
                \uElseIf{$c_m^{(j)}=\mathtt{advice}$}{
                    $r_m^{(j)}\leftarrow
                    \operatorname{AdviceAgent}
                    (t_m^{(j)},\mathcal{B}_u,\mathcal{P}_m^{(j)})$\;
                }
                \Else{
                    \textbf{raise} invalid task type\;
                }
            }

            $\mathcal{R}^{(j)}[\mathrm{id}_m^{(j)}]\leftarrow r_m^{(j)}$\;
        }

        $r^{(j)}\leftarrow r_{M_j}^{(j)}$; append
        $(i_j,r^{(j)})$ to $\mathcal{R}_{\mathcal{I}}$\;
    }

    $y\leftarrow\Gamma(q,\mathcal{R}_{\mathcal{I}})$\;
    \Return{$y$}\;
\end{algorithm}

\subsection{Query Agent}
\label{sec:query-agent}

\begin{figure}[htbp]
    \centering
    \includegraphics[width=\linewidth]{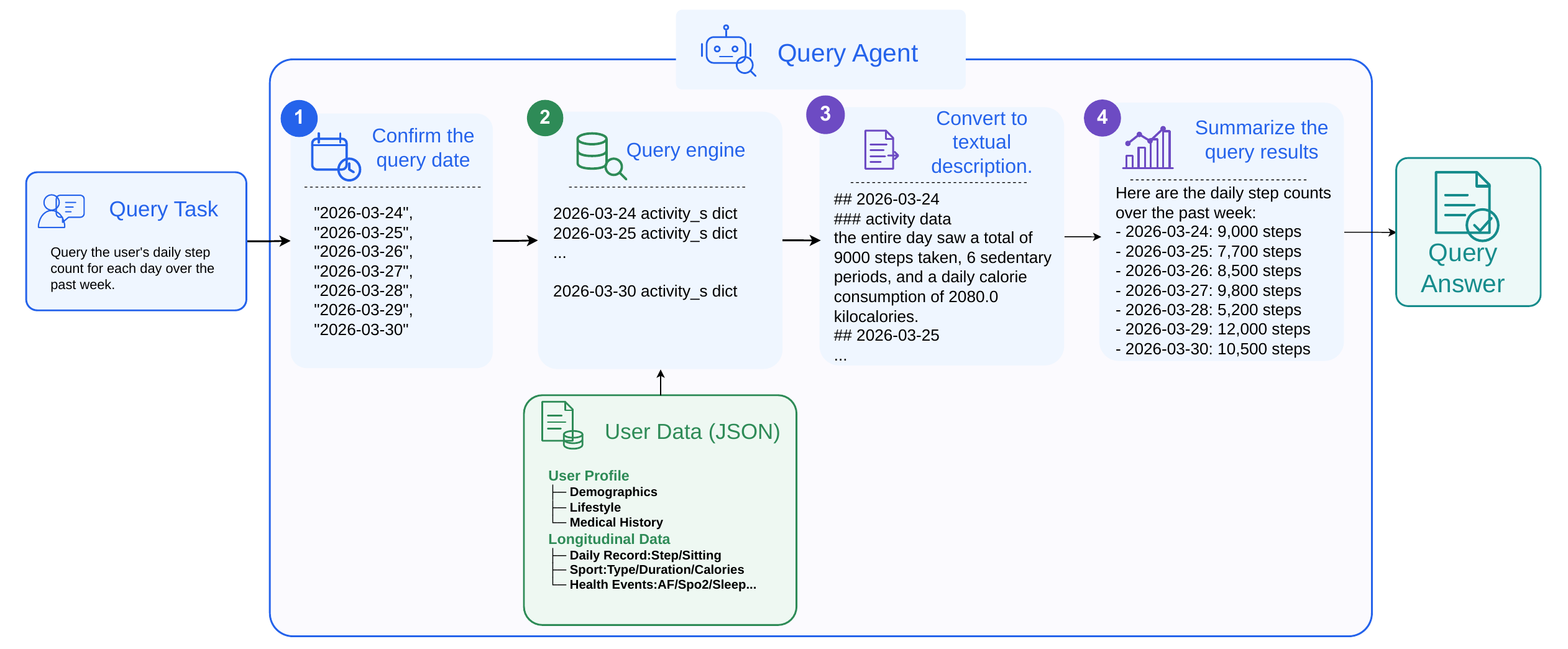}
    \caption{Data flow of the Query Agent. A natural-language temporal constraint is resolved into a date sequence; the corresponding structured records are retrieved and textualized to produce a traceable result.}
    \label{fig:query-agent}
\end{figure}

Figure~\ref{fig:query-agent} shows the Query Agent data flow from temporal resolution to
structured retrieval, textual evidence, and an answer.

Rather than prompting a single model to search the full textualized record, the Query Agent
separates temporal resolution, indexed access, evidence textualization, and answer
generation. For a complete Query Task, the pipeline is
\begin{equation}
\begin{aligned}
    \left(d_{\mathrm{ref}},\mathbf{d}_m^{(j)}\right)
    &=
    \operatorname{DateResolve}\left(t_m^{(j)}\right),\\
    \mathcal{X}_m^{(j)}
    &=
    \operatorname{QueryEngine}
    \left(\mathcal{D}_u,\mathbf{d}_m^{(j)}\right),\\
    \mathcal{S}_m^{(j)}
    &=
    \operatorname{Textualize}\left(\mathcal{X}_m^{(j)}\right),\\
    y_{\mathrm{Q},m}^{(j)}
    &=
    \operatorname{QueryAnswer}
    \left(t_m^{(j)},\mathcal{B}_u,\mathcal{S}_m^{(j)}\right),\\
    r_m^{(j)}
    &=
    \left(\mathcal{S}_m^{(j)},y_{\mathrm{Q},m}^{(j)}\right).
\end{aligned}
\label{eq:query-agent-pipeline}
\end{equation}
Here, DateResolve infers reference date $d_{\mathrm{ref}}$ and target dates
$\mathbf{d}_m^{(j)}$ from explicit, interval, or relative expressions in the task.
QueryEngine retrieves ordered records $\mathcal{X}_m^{(j)}$, and Textualize preserves
their dates, field semantics, and raw values in $\mathcal{S}_m^{(j)}$. QueryAnswer then
produces $y_{\mathrm{Q},m}^{(j)}$ from the task, profile, and evidence. Storing both
$\mathcal{S}_m^{(j)}$ and the answer in $r_m^{(j)}$ exposes traceable evidence to
explicitly dependent tasks without passing the complete record downstream.

\subsection{Analyse Agent}
\label{sec:analyse-agent}

Under the contract in Eq.~\eqref{eq:agent-interface-contracts}, the Analyse Agent
receives the complete Analyse Task $t_m^{(j)}$, the extracted profile $\mathcal{B}_u$,
and only the predecessor results in $\mathcal{P}_m^{(j)}$. Each predecessor result
retains its task identifier, task description, and returned evidence or answer, which
allows the agent to associate each finding with the retrieval operation that supports
it. The agent neither re-accesses $\mathcal{D}_u$ nor introduces values outside the
supplied results.

The agent supports two complementary analysis modes. For a single indicator, it orders
the supplied evidence by time and compares values or event frequencies across the
requested periods to characterize increases, decreases, stability, or fluctuations.
For a multi-indicator task, it organizes indicator-specific results under a shared
temporal scope and synthesizes their observed patterns without interpreting
co-occurrence as causal evidence. The processing flow is
\[
    \text{dependent results}
    \rightarrow
    \text{evidence organization}
    \rightarrow
    \text{temporal or multi-indicator comparison}
    \rightarrow
    \text{findings}.
\]

The output is an ordered sequence of evidence-grounded findings,
\[
    r_m^{(j)}
    =
    \left(
    f_{m,1}^{(j)},\ldots,f_{m,K_m}^{(j)}
    \right),
\]
where $K_m$ is the number of findings produced for task $t_m^{(j)}$. Each finding
identifies the relevant indicators and temporal scope, states the
observed comparison, and separates the supported conclusion from its interpretation
limits. If the supplied evidence is incomplete, the agent reports the limitation rather
than asserting a definitive trend. The result may become an explicit dependency of an
Advice Task.

\subsection{Advice Agent}
\label{sec:advice-agent}

Under Eq.~\eqref{eq:agent-interface-contracts}, the Advice Agent receives the complete
Advice Task $t_m^{(j)}$, $\mathcal{B}_u$, and selected Query or Analyse results in
$\mathcal{P}_m^{(j)}$. The profile supplies demographic, lifestyle, and health-history
context, whereas the dependency results constrain the recommendation to evidence
relevant to the current intent; results from unrelated intents are excluded.

The agent first identifies the observed evidence and the modifiable aspect addressed by
the task. It then connects the observation to a rationale and a practical action. When
supported by the evidence, the response also specifies an indicator to monitor or a
time horizon for reviewing the observed pattern. This process is summarized as
\[
    \begin{aligned}
    &\text{profile and dependent evidence}
    \rightarrow
    \text{evidence interpretation} \\
    &\qquad\rightarrow
    \text{personalized recommendation}
    \rightarrow
    \text{action and monitoring guidance}.
    \end{aligned}
\]
The output is an ordered recommendation sequence,
\[
    r_m^{(j)}
    =
    \left(
    a_{m,1}^{(j)},\ldots,a_{m,L_m}^{(j)}
    \right),
\]
where $L_m$ is the number of recommendations produced for task $t_m^{(j)}$. Each item
communicates a recommendation and its evidence-based rationale,
together with practical or monitoring guidance when supported.

The Advice Agent provides general health support rather than clinical diagnosis or
medical decision-making. It does not recommend changes to prescribed treatment, treats
wearable measurements as supportive rather than clinical evidence, and uses
conservative language when the available results are insufficient. Persistent or
concerning patterns may be accompanied by a suggestion to seek professional assessment;
the output does not replace professional judgment.

\subsection{Result Aggregation}
\label{sec:advice-aggregation}

For each intent $i_j$, the terminal task is constructed to produce an intent-level
result after consuming any explicitly required predecessors. A Query Intent terminates
in a Query result, whereas an Analyse or Advice Intent terminates in the corresponding
synthesized result. Intermediate task outputs remain in the intent-local
store and are exposed only through the dependency sets defined in
Eq.~\eqref{eq:dependency-results}.

Since $c_{M_j}^{(j)}=\kappa_j$, the Manager Agent selects only the terminal result
$r^{(j)}=r_{M_j}^{(j)}$ from each intent and preserves the recognized intent order:
\[
    \mathcal{R}_{\mathcal{I}}
    =
    \left(
    (i_1,r^{(1)}),
    (i_2,r^{(2)}),
    \ldots,
    (i_N,r^{(N)})
    \right).
\]
This order-preserving operation forms exactly one pair for each recognized intent; it
does not vote over agents, merge intermediate evidence across intents, or alter the
content of an agent result. Finally, $\Gamma$ in Eq.~\eqref{eq:response-aggregation}
renders $\mathcal{R}_{\mathcal{I}}$ in the context of the original query $q$, maintaining
the correspondence between each request and its response while preventing unrelated
intent-local evidence from entering the final context.

%% file: sections/synthetic_dataset.tex

\section{Synthetic Wearable Dataset Construction}
\label{sec:synthetic-dataset}

\begin{figure}[htbp]
	\centering
	\includegraphics[width=\linewidth]{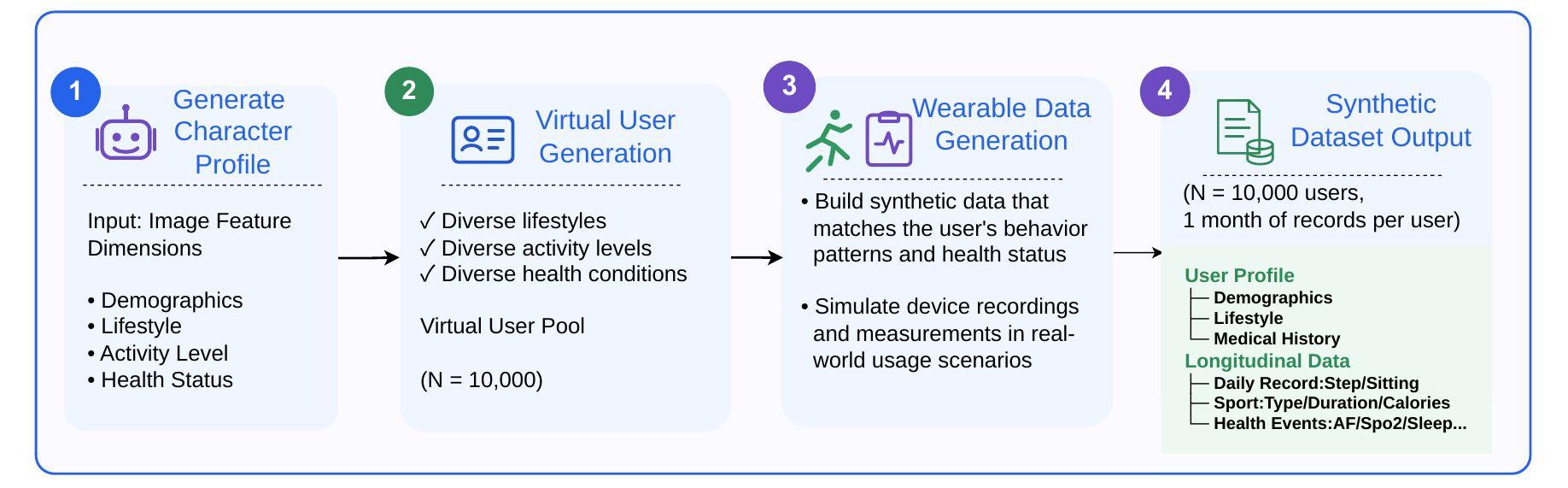}
	\caption{Generation pipeline for the synthetic wearable health dataset. Virtual user profiles condition the generation of longitudinal wearable records, which are then organized into the evaluation dataset.}
	\label{fig:data-synthesis}
\end{figure}

Large-scale real-world wearable datasets with consistent longitudinal structure and
annotations are difficult to obtain and raise substantial privacy concerns. We therefore construct a synthetic
evaluation dataset through the two-stage pipeline in Figure~\ref{fig:data-synthesis}:
diverse virtual profiles are generated first and then condition longitudinal records.
The dataset is not intended to reproduce disease prevalence, physiological
distributions, or clinical progression, and supports no clinical or epidemiological
inference.

\subsection{Profile-Conditioned Data Generation}
\label{subsec:synthetic-generation-pipeline}

The first stage generates $10{,}000$ profiles containing demographic, lifestyle, and
health-history information. Each profile conditions the subsequent record generation.

The second stage generates records for March 1--30, 2026. Dates for the same user share
one profile context, whereas different users retain their respective conditions.

At runtime, stable profile information and time-varying date-indexed records form
$\mathcal{D}_u$, the Query Agent's data source. The \texttt{role} field of $\mathcal{D}_u$ corresponds to
the extracted profile $\mathcal{B}_u$, and \texttt{data\_s} stores the daily records.

\subsection{Longitudinal Wearable Data Structure}
\label{subsec:synthetic-record-structure}

The hierarchical record $\mathcal{D}_u$ combines user-level demographics, lifestyle,
and health history with daily entries indexed by \texttt{datetime}. Each entry contains
\texttt{sport\_s}, \texttt{health\_s}, and \texttt{activity\_s};
Table~\ref{tab:wearable-schema} summarizes their evaluation uses.

\begin{table}[htbp]
	\centering
	\caption{Primary components of the synthetic wearable dataset and their evaluation uses.}
	\label{tab:wearable-schema}
	\small
	\setlength{\tabcolsep}{5pt}
	\begin{tabular}{L{2.7cm} L{5.2cm} L{4.2cm}}
		\toprule
		\textbf{Category} &
		\textbf{Content} &
		\textbf{Supported Tasks} \\
		\midrule
		
		User profile &
		Demographics, lifestyle, and health history &
		Personalized query construction and user-context conditioning \\
		
		Exercise records &
		Exercise time, type, duration, and calories burned &
		Exercise retrieval and longitudinal change analysis \\
		
		Health monitoring &
		Atrial fibrillation, premature heartbeat events, blood oxygen, and sleep apnea &
		Health-event retrieval and temporal trend analysis \\
		
		Sleep information &
		Sleep timing, duration, and quality score &
		Sleep-duration and sleep-quality change analysis \\
		
		Daily activity &
		Daily step count, sedentary episodes, and total calorie expenditure &
		Daily activity retrieval and longitudinal trend analysis \\
		
		\bottomrule
	\end{tabular}
\end{table}

The \texttt{sleep\_stage} field contains sleep onset, end time, duration, and quality
score, but not fine-grained REM, deep, or light stages; evaluation therefore uses only
sleep duration and quality. Records follow the Query Agent's structured JSON format.
Appendix~\ref{app:json-schema} shows a simplified structure, and the code provides the
complete schema.

\subsection{Scale, Quality Control, and Evaluation Use}
\label{subsec:synthetic-dataset-statistics}

The monthly records cover exercise, health monitoring, sleep, and daily activity,
supporting questions with varied temporal ranges, indicator combinations, and
complexity.

Pydantic models constrain generation; the pipeline then requires \texttt{data\_s} and
at least 20 daily records, otherwise regenerating records from the original profile.
These checks cover only basic structure and count, not date continuity, field ranges, or
physiological consistency (Section~\ref{subsec:limitations}).

As a \emph{synthetic evaluation dataset}, it supports structured retrieval,
multi-intent recognition, and multidimensional longitudinal analysis. Section~\ref{sec:experiments}
specifies sampling and metrics; Appendix~\ref{app:synthetic-data-details} provides
generation prompts and quality-control details.

%% file: sections/experiments.tex

\section{Experiments}
\label{sec:experiments}

\subsection{Experimental Setup}
\label{subsec:experimental-setup}

All experiments use the $10{,}000$-user synthetic dataset in
Section~\ref{sec:synthetic-dataset}; each subsection specifies its sample and questions.

All evaluated methods use \texttt{deepseek-v4-pro} at temperature $1.0$. The Direct LLM
baseline generates an answer from the query and the textualized complete record; our
method performs intent recognition, decomposition, and specialized execution with the
same model and inference configuration.

For overall response quality, an independent invocation of \texttt{deepseek-v4-pro} at
temperature $1.0$ scores Trustworthiness, Transparency, and Actionability as specified
in Section~\ref{subsec:overall-quality}. Each answer is scored once and independently;
the evaluator receives no method identity, and category means are computed over all
instances in each question category.

\begin{figure}[!t]
	\centering
	\includegraphics[width=\linewidth]{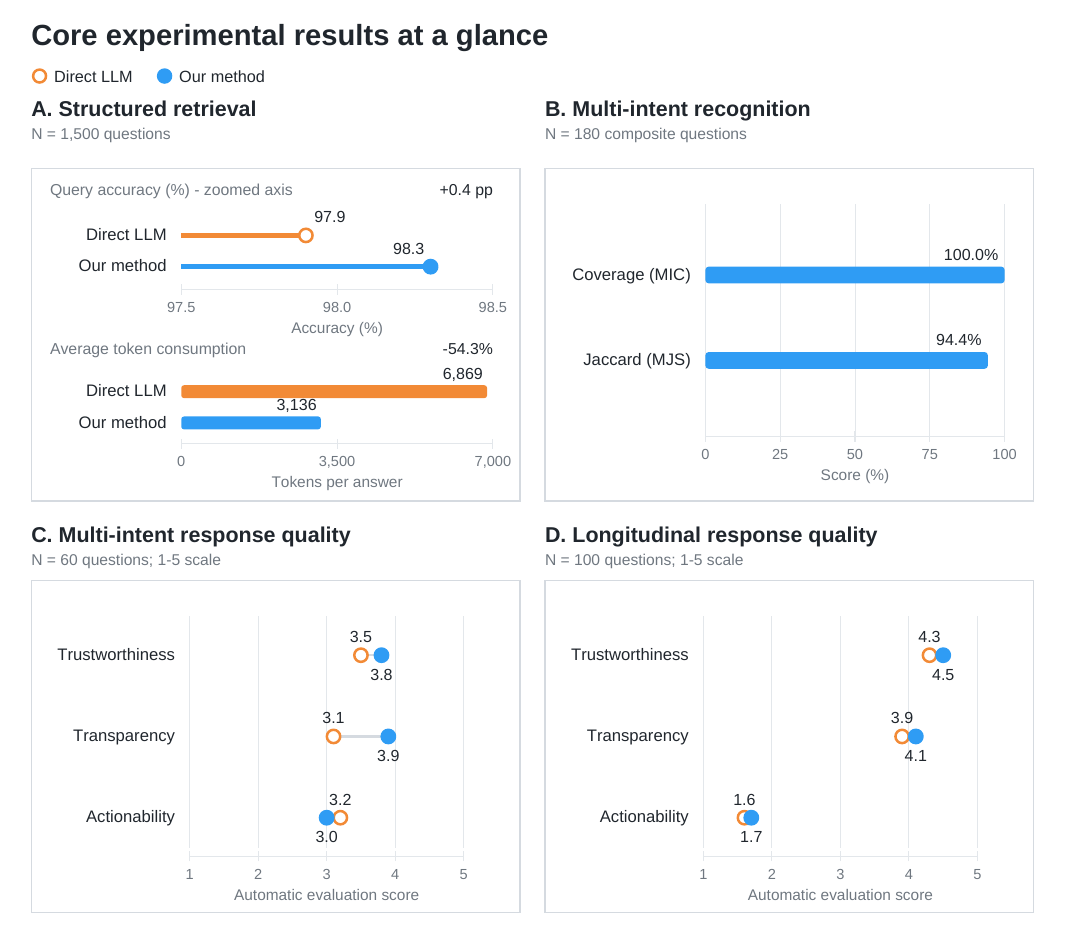}
	\caption{Overview of the main experimental results. Panel~(a) shows that the Query
	Agent achieves comparable exact-value accuracy to Direct LLM while using fewer
	query-stage tokens; its accuracy axis is explicitly zoomed to $97.5$--$98.5\%$.
	Panel~(b) reports absolute Multi-Intent Coverage and Multiset Jaccard scores without
	an external structured-output baseline. Panels~(c)--(d) report mean automatic
	Trustworthiness, Transparency, and Actionability scores on a $1$--$5$ scale. The
	figure presents observed averages and does not imply statistical significance.}
	\label{fig:results-overview}
\end{figure}

Figure~\ref{fig:results-overview} summarizes the three evaluation dimensions before the
detailed protocols and metrics are introduced. Panel~(a) compares structured-retrieval
accuracy and query-stage token consumption; Panel~(b) reports the Manager Agent's
absolute intent-recognition performance; and Panels~(c)--(d) compare automatic
response-quality scores on multi-intent and longitudinal questions. The figure provides
a cross-experiment overview, while the following subsections and tables report the
corresponding benchmark construction, metric definitions, and exact results.

\subsection{Query Agent Evaluation}
\label{subsec:query-agent-eval}

\subsubsection{Experimental Design}
\label{subsubsec:query-agent-design}

The retrieval benchmark covers exercise count, atrial fibrillation, premature heartbeat events,
sleep apnea, daily steps, calorie expenditure, and sedentary episodes. It uses
single-day, continuous-window, and multi-period retrieval.

For each of $300$ randomly sampled users, we select five topics, assign a temporal form,
and personalize the wording with the user profile, yielding $1{,}500$ instances.
References are derived from the structured records and manually reviewed.

Both methods receive the same question and record. Direct LLM receives the textualized
complete record, whereas the Query Agent retrieves target records through
\texttt{DateResolve} and \texttt{QueryEngine} before \texttt{QueryAnswer}; the comparison
therefore isolates data-access and context-construction strategies.

\subsubsection{Metrics}
\label{subsubsec:query-agent-metrics}

We evaluate retrieval correctness and token use.

\paragraph{Query Accuracy}

For $N_{\mathrm{Q}}$ instances, let $\hat{y}_n$ and $y_n^{*}$ be the system and
reference answers, and let $\nu(\cdot)$ extract the ordered target value sequence. Exact
value equality defines
\begin{equation}
	\operatorname{Judge}\left(\hat{y}_n,y_n^{*}\right)
	=
	\mathbb{I}\!\left[
		\nu\!\left(\hat{y}_n\right)
		=
		\nu\!\left(y_n^{*}\right)
	\right],
	\label{eq:query-judge}
\end{equation}
where $\nu:\mathcal{Y}\rightarrow\mathcal{V}^{*}$ extracts an ordered sequence of target
values from an answer in the answer space $\mathcal{Y}$, and $\mathbb{I}[\cdot]$ is the
indicator function. Thus, Eq.~\eqref{eq:query-judge} ignores wording differences but
requires the extracted numerical values and their order to match exactly. The overall
metric is
\begin{equation}
	\mathrm{Query\ Accuracy}
	=
	\frac{1}{N_{\mathrm{Q}}}
	\sum_{n=1}^{N_{\mathrm{Q}}}
	\operatorname{Judge}\left(\hat{y}_n,y_n^{*}\right).
	\label{eq:query-accuracy}
\end{equation}
Eq.~\eqref{eq:query-accuracy} averages the exact-value judgment over all retrieval
instances.

\paragraph{Token Consumption}

Average recorded token consumption characterizes model context under each access
strategy. Because the pipelines invoke the model differently, it is a query-stage
measure rather than end-to-end efficiency.

\subsubsection{Results}
\label{subsubsec:query-agent-results}

\input{tables/main_results}

Across $1{,}500$ instances, Table~\ref{tab:query-results} reports $97.9\%$ accuracy for
Direct LLM and $98.3\%$ for the Query Agent; the $0.4$-point difference is not tested
for significance. Average token use is $6{,}869$ versus $3{,}136$, indicating less
query-stage context for the Query Agent, not lower end-to-end system cost.

\subsection{Manager Agent Multi-Intent Recognition Evaluation}
\label{subsec:multi-intent-eval}

\subsubsection{Experimental Design}
\label{subsubsec:multi-intent-design}

The benchmark tests whether the Manager Agent recovers distinct intents, including
repeated types, from $\mathcal{C}$ in Section~\ref{sec:problem-formulation}.

We define six representative intent-type combinations:
\begin{center}
\small
\texttt{query\&query}, \texttt{query\&analyse}, \texttt{query\&advice},\\
\texttt{query\&query\&query}, \texttt{query\&query\&advice}, and
\texttt{query\&analyse\&advice}.
\end{center}
For each of $30$ sampled users, an LLM generates all six combinations conditioned on the
profile, yielding $180$ manually reviewed questions with annotated intent references.

Only the predicted and reference intent-type lists are compared; decomposition,
dependency construction, and specialized execution are outside this experiment.

\subsubsection{Metrics}
\label{subsubsec:multi-intent-metrics}

We use Multi-Intent Coverage (MIC) for type coverage and Multiset Jaccard Similarity for
repeated-type counts.

Let the benchmark contain $M\geq 1$ instances. For instance $n$, let the predicted intent-type
list and manually annotated reference list be
\[
P_n=(p_{n,1},\ldots,p_{n,\hat N_n}),
\qquad
G_n=(g_{n,1},\ldots,g_{n,N_n}),
\]
where $\hat N_n$ and $N_n\geq 1$ are the predicted and reference lengths.

\paragraph{Multi-Intent Coverage (MIC)}

Following \citet{godbole2004discriminative}, let
$\bar P_n=\operatorname{set}(P_n)$ and $\bar G_n=\operatorname{set}(G_n)$. Instance and
macro MIC are
\begin{align}
	\mathrm{MIC}_n
	=
	\frac{
		\left|
		\bar{P}_n \cap \bar{G}_n
		\right|
	}{
		\left|
		\bar{G}_n
		\right|
	},
	\label{eq:instance-mic}
	\\
	\mathrm{MIC}
	=
	\frac{1}{M}
	\sum_{n=1}^{M}
	\mathrm{MIC}_n.
	\label{eq:mic}
\end{align}

Eq.~\eqref{eq:instance-mic} measures per-instance type coverage, and
Eq.~\eqref{eq:mic} reports its macro average. Because set conversion removes duplicates,
MIC does not assess multiplicity.

\paragraph{Multiset Jaccard Similarity}

Let $c_{n,k}^{\mathrm{pred}}$ and $c_{n,k}^{\mathrm{gold}}$ count type
$k\in\mathcal{C}$ in $P_n$ and $G_n$. Multiset intersection and union use minimum and
maximum counts~\citep{ngo2021structural}:
\begin{align}
	J_n
	=
	\frac{
		\displaystyle
		\sum_{k\in\mathcal{C}}
		\min
		\left(
		c_{n,k}^{\mathrm{pred}},
		c_{n,k}^{\mathrm{gold}}
		\right)
	}{
		\displaystyle
		\sum_{k\in\mathcal{C}}
		\max
		\left(
		c_{n,k}^{\mathrm{pred}},
		c_{n,k}^{\mathrm{gold}}
		\right)
	},
	\label{eq:instance-multiset-jaccard}
	\\
	J
	=
	\frac{1}{M}
	\sum_{n=1}^{M}
	J_n.
	\label{eq:multiset-jaccard}
\end{align}

Eq.~\eqref{eq:instance-multiset-jaccard} preserves per-instance type multiplicity, and
Eq.~\eqref{eq:multiset-jaccard} reports its macro average. Both metrics range from $0$
to $1$. For reference
\texttt{[query, query, advice]} and prediction \texttt{[query, advice]}, MIC is $1$
but Multiset Jaccard is $2/3$; neither metric assesses intent order.

\subsubsection{Results}
\label{subsubsec:multi-intent-results}

\input{tables/intent_results}

Table~\ref{tab:intent-results} reports $100.0\%$ MIC and $94.4\%$ Multiset Jaccard
across $180$ questions. Without an external baseline producing the same structure,
these scores characterize absolute benchmark performance.

\subsection{Overall Response-Quality Evaluation}
\label{subsec:overall-quality}

\subsubsection{Experimental Design}
\label{subsubsec:overall-quality-design}

We evaluate final responses on multi-intent questions, which require coverage of
distinct requests, and multidimensional longitudinal-trend questions, which require
comparison of several indicators over time.

For $10$ sampled users, all six intent combinations yield $60$ multi-intent instances;
the same $10$ manually designed trend questions are instantiated for each user, spanning
exercise, daily activity, sleep, and health monitoring and yielding $100$ more instances.
All $160$ question instances are manually checked to ensure that their temporal ranges
and target indicators are supported by the corresponding records.

Both methods use the same query, record, model, and automatic protocol. The evaluator
sees only the query and final response, not method identity or intermediate state.
Appendix~\ref{app:quality-rubric} provides the rubric and protocol.

\subsubsection{Metrics}
\label{subsubsec:overall-quality-metrics}
An LLM assigns separate 1--5 scores for Trustworthiness (evidence and its link to
conclusions), Transparency (clarity from data to conclusion), and Actionability
(specific and useful advice); we report category means without combining dimensions.

Because the evaluator lacks the structured record, Trustworthiness measures evidence
presentation and evidence--conclusion linkage, not independent factual consistency;
Query Accuracy evaluates numerical retrieval separately. Actionability on non-advice
questions is interpreted relative to the task objective.

\subsubsection{Results}
\label{subsubsec:overall-quality-results}

\input{tables/reasoning_results}

On multi-intent questions, Table~\ref{tab:multi-intent-quality} reports respective
Trustworthiness, Transparency, and Actionability scores of $3.8$, $3.9$, and $3.0$ for
our method, versus $3.5$, $3.1$, and $3.2$ for Direct LLM. On longitudinal trends,
Table~\ref{tab:longitudinal-quality} reports $4.5$, $4.1$, and $1.7$ versus $4.3$,
$3.9$, and $1.6$.
Trustworthiness and Transparency are higher for our method in both categories, whereas
Actionability does not improve consistently.

%% file: tables/main_results.tex
\begin{table}[htbp]
	\centering
	\caption{Structured wearable-data retrieval results on $N=1{,}500$ instances. Both methods use the same foundation model and underlying records. Token statistics cover only the current retrieval pipeline and do not represent end-to-end system cost.}
	\label{tab:query-results}
	\begin{tabular}{lcc}
		\toprule
		\textbf{Method} &
		\textbf{Query Accuracy} &
		\textbf{Average Token Consumption} \\
		\midrule
		Direct LLM &
		97.9\% &
		6,869 \\
		Query Agent &
		\textbf{98.3\%} &
		\textbf{3,136} \\
		\bottomrule
	\end{tabular}
\end{table}

%% file: tables/intent_results.tex
\begin{table}[htbp]
	\centering
	\caption{Multi-intent recognition results for the Manager Agent on $N=180$ questions. MIC measures intent-type coverage, while Multiset Jaccard Similarity preserves repeated-intent counts.}
	\label{tab:intent-results}
	\begin{tabular}{lcc}
		\toprule
		\textbf{Method} &
		\textbf{MIC} &
		\textbf{Multiset Jaccard Similarity} \\
		\midrule
		Manager Agent &
		\textbf{100.0\%} &
		\textbf{94.4\%} \\
		\bottomrule
	\end{tabular}
\end{table}

%% file: tables/reasoning_results.tex
\begin{table}[htbp]
	\centering
	\caption{Automatic overall response-quality evaluation on $N=60$ multi-intent questions (1--5; higher is better).}
	\label{tab:multi-intent-quality}
	\begin{tabular}{lccc}
		\toprule
		Method &
		Trustworthiness &
		Transparency &
		Actionability \\
		\midrule
		Direct LLM &
		3.5 &
		3.1 &
		\textbf{3.2} \\
		Our method &
		\textbf{3.8} &
		\textbf{3.9} &
		3.0 \\
		\bottomrule
	\end{tabular}
\end{table}

\begin{table}[htbp]
	\centering
	\caption{Automatic overall response-quality evaluation on $N=100$ multidimensional longitudinal-trend questions (1--5; higher is better).}
	\label{tab:longitudinal-quality}
	\begin{tabular}{lccc}
		\toprule
		Method &
		Trustworthiness &
		Transparency &
		Actionability \\
		\midrule
		Direct LLM &
		4.3 &
		3.9 &
		1.6 \\
		Our method &
		\textbf{4.5} &
		\textbf{4.1} &
		\textbf{1.7} \\
		\bottomrule
	\end{tabular}
\end{table}

%% file: sections/results_discussion.tex

\section{Discussion}
\label{sec:results-discussion}

We interpret the results through structured data access, multi-intent organization,
and final response quality. Because the evaluation lacks complete component-level
ablations, the observed patterns characterize the framework rather than the effects of
individual modules.

\subsection{Structured Data Access and Context Construction}
\label{subsec:discussion-query}

Both methods achieve high retrieval accuracy but construct the model context
differently. Direct LLM jointly resolves time, locates data, and generates an answer
from the complete record, whereas the Query Agent resolves dates before accessing
target records through temporal indices and structured fields. Its lower token use in
Table~\ref{tab:query-results} therefore reflects task-relevant query-stage context
construction, not stronger language-model reasoning or lower end-to-end cost. Because
the accuracy difference is small and has not been tested for statistical significance,
the evidence supports only comparable accuracy with less retrieval context.

\subsection{Multi-Intent Recognition and Task Organization}
\label{subsec:discussion-multi-intent}

The gap between Multi-Intent Coverage and Multiset Jaccard Similarity in
Table~\ref{tab:intent-results} indicates reliable type coverage but remaining errors in
the multiplicity of repeated intent types. Neither metric evaluates intent boundaries,
textual spans, or relative order; intent-level Exact Match, order consistency, and
content matching would provide a finer-grained assessment.

After recognition, explicit same-intent dependencies make the execution structure
inspectable and restrict downstream contexts to selected predecessor results. However,
the recognition experiment ends before task decomposition, and no ablation isolates
dependency construction. Its contribution therefore remains a design property rather
than an independently verified source of performance gain.

\subsection{Final Response Quality and Health Advice}
\label{subsec:discussion-response-quality}

The higher mean Trustworthiness and Transparency scores in
Tables~\ref{tab:multi-intent-quality} and~\ref{tab:longitudinal-quality} are consistent
with the framework's explicit organization of requests, evidence, and analytical
outputs. In particular, the larger Transparency difference on multi-intent questions
suggests that intent-level organization can help connect indicator-specific evidence to
the corresponding explanations. This interpretation is not causal because the current
evaluation does not isolate decomposition, specialized execution, or aggregation.

The evaluator sees only the user query and final response. Trustworthiness thus measures
evidence presentation and evidence--conclusion linkage rather than independent factual
consistency with the structured record; underlying-value access is assessed separately
by Query Accuracy.

Actionability does not improve consistently. Potential factors include the Advice
Agent's lack of explicit modeling for user preferences, practical constraints, and
longitudinal feedback, as well as the inclusion of trend questions that do not request
an action. Future evaluation should report Actionability separately for questions with
an \texttt{advice} intent and assess advice for evidence consistency, specificity,
personalization, and executability.

\subsection{Limitations}
\label{subsec:limitations}

The synthetic monthly records provide controlled structured inputs but may not reflect
real user distributions, long-term variation, or physiological relationships. Current
quality control verifies the top-level structure and a minimum record count, not complete
date continuity, field ranges, or cross-indicator consistency. The experiments therefore
evaluate task organization and data processing rather than clinical effectiveness or
performance over quarterly and annual records.

The comparison scope is also limited. Multi-intent recognition lacks an external
structured-output baseline, the framework has not undergone component-level ablation,
and all methods use one foundation model. Consequently, the results do not separately
quantify the contributions of recognition, dependency construction, or individual
agents, nor establish cross-model generalization.

Finally, overall response quality relies on LLM-based evaluation, whose preferences and
scoring variability may affect the results. The evaluator cannot independently verify
the underlying health values, and human expert assessment remains necessary. Efficiency
is likewise limited to query-stage token use rather than model invocations, end-to-end
latency, or cost. Future work should address these limitations with real or de-identified
long-term data, component ablations, cross-model experiments, expert evaluation, and
complete system-level efficiency measurements.

%% file: sections/conclusion.tex

\section{Conclusion}
\label{sec:conclusion}

We presented a task-oriented multi-agent framework for complex longitudinal health
queries over wearable data. The framework organizes execution around user intents and
executable tasks. A Manager Agent performs multi-intent recognition, fine-grained task
decomposition, and explicit dependency construction, then invokes the Query, Analyse,
or Advice Agent according to task type for structured retrieval, longitudinal
multi-indicator analysis, or health advice generation. Intents maintain isolated task
structures and local execution states before the Manager Agent aggregates their results.

On the synthetic wearable health dataset, the framework supports structured retrieval,
multi-intent recognition, and evidence-organized final responses. The results indicate
that task-relevant record access can reduce query-stage context while maintaining high
retrieval accuracy, and that explicit intent organization can preserve repeated request
types. The response-quality results favor our method in mean Trustworthiness and
Transparency but not consistently in Actionability.

The current evidence is limited to monthly synthetic data and automatic evaluation and
lacks a structured multi-intent baseline and component ablations. Future work will use
real long-term data, cross-model experiments, human expert evaluation, and end-to-end
efficiency analysis to examine generalization and isolate the contribution of each
component.

%% file: sections/appendix.tex

\section{Automatic Overall Response-Quality Evaluation}
\label{app:quality-rubric}
\label{app:automatic-evaluation}

\subsection{Scoring Rubric}

Automatic evaluation scores Trustworthiness, Transparency, and Actionability on the
following 1--5 rubric.

\input{tables/quality_evaluation_rubric}

\subsection{Evaluation Protocol}
\label{app:judge-prompt}

For each instance, the evaluator receives the user question and one final answer, applies
the rubric in Table~\ref{tab:quality_evaluation_rubric}, and returns the three integer scores together
with a brief rationale in JSON. The evaluator is not given the method identity,
intermediate task results, or the underlying structured record. The exact executable
prompt is released with the code at
\url{https://github.com/yangkunpeng-coder/WearableDeviceAgents}.

\section{Multi-Agent System Prompts}
\label{app:agent-prompts}

This appendix summarizes the prompt interfaces used by the main LLM components. The
exact executable prompts and Pydantic schemas are released with the open-source code.
The implementation and the paper consistently use \texttt{analyse} as the machine-readable
type value and \emph{Analyse Agent} in prose. The \texttt{base\_info} field is
deterministically parsed from the \texttt{role} field in $\mathcal{D}_u$ and corresponds
to $\mathcal{B}_u$.

\paragraph{Manager Agent.}
\label{app:prompt-intent-recognition}
The recognition prompt maps a question to an ordered list of intent types and
self-contained descriptions while preserving shared constraints. The decomposition
prompt maps each intent to an ordered task list containing identifiers, types,
descriptions, and same-intent predecessor dependencies.
\label{app:prompt-task-decomposition}

\paragraph{Query Agent.}
\label{app:prompt-query-agent}
\label{app:prompt-query-date}
\label{app:prompt-query-answer}
The date-resolution prompt infers the reference date and target dates from the complete
Query Task. After
deterministic record access and textualization, the answer prompt maps the complete task,
user profile, and retrieved evidence to a natural-language task result.

\paragraph{Analyse and Advice Agents.}
\label{app:prompt-analyse-agent}
\label{app:prompt-advice-agent}
Both prompts receive the complete task, user profile, and explicitly selected predecessor
results. The Analyse Agent organizes longitudinal or multi-indicator evidence; the
Advice Agent produces general health-support recommendations grounded in that evidence.

\paragraph{Result aggregation.}
\label{app:prompt-aggregation}
The aggregation prompt receives the original question and ordered intent-level results,
then produces one response that covers every recognized request without accessing the
underlying record again.

\section{System Implementation Details}
\label{app:implementation-details}

This section delineates the implementation boundaries among large language models, deterministic procedures, and structured data access. LLMs perform semantic interpretation and text generation; program logic handles agent selection, dependency resolution, and state management; and structured interfaces directly retrieve the underlying records.

\subsection{System Components and Implementation Boundaries}
\label{app:component-implementation}

Table~\ref{tab:component-implementation} summarizes the inputs, outputs, and implementation of the principal components. The system deterministically selects a specialized agent according to \texttt{task\_type}.

\begingroup
	\small
	\setlength{\tabcolsep}{3.5pt}
	\begin{longtable}{
			L{3.0cm}
			L{3.0cm}
			L{4.2cm}
			L{4.3cm}
		}
		\caption{Principal components of the multi-agent system and their implementations.}
		\label{tab:component-implementation} \\
		\toprule
		\textbf{Component} &
		\textbf{Implementation} &
		\textbf{Input} &
		\textbf{Output} \\
		\midrule
		\endfirsthead
		\caption[]{Principal components of the multi-agent system and their implementations (continued).} \\
		\toprule
		\textbf{Component} &
		\textbf{Implementation} &
		\textbf{Input} &
		\textbf{Output} \\
		\midrule
		\endhead
		\midrule
		\multicolumn{4}{r}{\small Continued on next page} \\
		\endfoot
		\bottomrule
		\endlastfoot

		\texttt{ExtractProfile} &
		Deterministic &
		Complete structured user record $\mathcal{D}_u$ &
		User profile $\mathcal{B}_u$ \\
		
		\texttt{RecognizeIntents} &
		LLM &
		Original user question &
		Intent types and self-contained descriptions ordered as in the original question \\
		
		\texttt{DecomposeTasks} &
		LLM &
		Current intent type and description &
		Ordered tasks with identifiers, types, descriptions, and explicit dependencies \\
		
		Agent Selection &
		Deterministic &
		Current task's \texttt{task\_type} &
		Query Agent, Analyse Agent, or Advice Agent \\
		
		\texttt{CollectResults} &
		Deterministic &
		Current task's \texttt{dependencies} and local results for the current intent &
		Predecessor results on which the current task actually depends \\
		
		\texttt{DateResolve} &
		LLM &
		Complete Query Task &
		Inferred reference date and target date sequence \\
		
		\texttt{QueryEngine} &
		Structured Data Access &
		Target date sequence and structured wearable record &
		Raw structured records for the target dates \\
		
		\texttt{Textualize} &
		Deterministic &
		Structured records returned by the Query Engine &
		Textual evidence retaining dates, field semantics, and raw values \\
		
		\texttt{QueryAnswer} &
		LLM &
		Query Task, user profile, and textualized evidence &
		Natural-language answer to the current Query Task \\
		
		\texttt{AnalyseAgent} &
		LLM &
		Analyse Task, user profile, and explicit dependency results &
		Longitudinal comparison or multi-indicator analysis \\
		
		\texttt{AdviceAgent} &
		LLM &
		Advice Task, user profile, and explicit dependency results &
		Health-support recommendations based on available health information \\
		
		\texttt{Aggregate} &
		LLM &
		Original user question and intent-level results in the original intent order &
		Complete user-facing answer \\
		
	\end{longtable}
\endgroup

Components marked as LLM perform natural-language understanding or generation. The remaining components use deterministic procedures or structured data access; the LLM does not generate the underlying health measurements.

\subsection{Runtime Data Flow}
\label{app:dependency-state-management}
\label{app:query-agent-implementation}
\label{app:specialized-agent-input}
\label{app:result-management}

Each intent maintains an intent-local task sequence and result store. A Query Task has no predecessors. Analyse and Advice Tasks receive only same-intent predecessor results explicitly listed in \texttt{dependencies} through \texttt{CollectResults}; unspecified results and state from other intents are excluded from their context.

The Query Agent follows Eq.~\eqref{eq:query-agent-pipeline}: \texttt{Date\allowbreak Resolve} and \texttt{Query\allowbreak Answer} are LLM-based, whereas \texttt{Query\allowbreak Engine} and \texttt{Textualize} are deterministic. Each task result is stored under its \texttt{task\_id}, and the result of a terminal task becomes the intent-level result. Once all intents have been processed, the Manager Agent aggregates these results in the order of the original question without accessing the underlying records again.

\section{Synthetic Wearable Data Generation Details}
\label{app:synthetic-data-details}

This appendix describes profile generation, longitudinal record generation, the data structure, and quality checks for the synthetic dataset. The complete JSON Schema is defined in code using Pydantic.

\subsection{Synthetic User Profile Generation}
\label{app:profile-generation}

Synthetic user profiles are first generated by an LLM and then used to condition longitudinal wearable-record generation. The implementation uses \texttt{deepseek-chat} at temperature $1.5$; this data-generation model is distinct from the \texttt{deepseek-v4-pro} model used in the experiments. Each profile contains demographic information, lifestyle information, and medical history. The generation instructions request a relatively balanced distribution across age groups and require approximately 50\% of users to have at least one of atrial fibrillation, premature heartbeats, sleep apnea, or insufficient sleep duration.

The exact executable profile-generation prompt is provided in the open-source code.

Each synthetic user profile contains the following fields:

\begin{itemize}
	\item \texttt{name\_id}: unique identifier of the synthetic user;
	\item \texttt{demographic\_information}: demographic information;
	\item \texttt{lifestyle}: lifestyle information; and
	\item \texttt{medical\_history}: medical history.
\end{itemize}

The generation script targets 10,000 users and creates profiles in batches. After all batches have completed, their outputs are merged and \texttt{name\_id} values are reassigned sequentially.

\subsection{Longitudinal Wearable-Record Generation}
\label{app:wearable-generation}

After profile generation, the system generates longitudinal wearable records conditioned
on the user's demographic, lifestyle, and medical-history fields. The implementation
uses \texttt{deepseek-chat} at temperature $1.5$. The
experimental code sets the record-generation interval to 2026-03-01 through 2026-03-30.

The exact executable wearable-record generation prompt is provided in the open-source
code.

Here, \texttt{day\_num} is calculated from the interval between the start and end dates;
\texttt{start\_\allowbreak datetime} and \texttt{end\_\allowbreak datetime}
denote the start and end dates of record generation, respectively. Demographic information, lifestyle, and medical history from the user profile are supplied directly as prompt conditions so that the generated longitudinal records reflect the background of each synthetic user.

The generated output is constrained by the Pydantic-defined \texttt{DataListForm}; each date is associated with exercise, health-monitoring, and daily-activity records.

\subsection{Data Structure and JSON Schema}
\label{app:json-schema}

The synthetic data use the date as the basic record unit, and each record contains \texttt{sport\_s}, \texttt{health\_s}, and \texttt{activity\_s}. A simplified JSON structure is shown below; the complete object definitions are available in the open-source code.

\begin{lstlisting}[
	style=promptstyle,
	caption={Simplified JSON structure of the synthetic wearable data.},
	label={lst:wearable-json-schema}
	]
	{
		"role": {
			"name_id": "<integer>",
			"demographic_information": "<string>",
			"lifestyle": "<string>",
			"medical_history": "<string>"
		},
		"data_s": [
			{
				"datetime": "YYYY-MM-DD",
				"sport_s": ["<exercise records>"],
				"health_s": ["<health and sleep records>"],
				"activity_s": ["<daily activity records>"]
			}
		]
	}
\end{lstlisting}

The complete JSON object corresponds to the structured user record $\mathcal{D}_u$. The
\texttt{role} field represents the user profile, whereas \texttt{data\_s} contains the
longitudinal wearable records organized by date. Their relationship to the notation in the main text is
\[
    \mathcal{D}_u
    \leftrightarrow
    \{\texttt{role},\texttt{data\_s}\},
    \qquad
    \mathcal{B}_u
    \leftrightarrow
    \texttt{role}.
\]

Exercise records contain start and end times, exercise type, duration, and calories. Health records contain fields for atrial fibrillation, premature heartbeat events, blood oxygen, sleep apnea, and sleep. Activity records contain step count, sedentary-event count, and total calories. The open-source schema provides the complete field types and constraints.

\subsection{Data Quality Checks}
\label{app:data-validation}

During generation, Pydantic data models first constrain the LLM's structured output. A longitudinal record is saved only if the generated object contains a \texttt{data\_s} field, preventing outputs that do not conform to the target top-level structure from being written directly to the final data file.

After generation, the code performs a record-completeness check. It first verifies the presence of \texttt{data\_s}; users without this field are marked as having a format error. It then counts the dated records in \texttt{data\_s}; users with fewer than 20 records are marked as having an abnormal record count.

For users with too few dated records, the code repeats the generation process using the original user profile and the same longitudinal data-generation procedure.

The current post-processing checks cover only the top-level JSON structure and the number of dated records. Section~\ref{subsec:limitations} discusses the resulting data-quality limitations.

\section{Evaluation Dataset Examples}
\label{app:evaluation-examples}

This appendix provides representative questions from the evaluation tasks to illustrate their formats and complexity.

\subsection{Structured-Data Query Examples}
\label{app:query-examples}

Table~\ref{tab:query-examples} presents representative questions from the structured-data query evaluation set. Each question is associated with the structured wearable records of one synthetic user and requires access to data for the specified time range and health indicators. The set includes single-date queries, continuous-range queries, joint queries over noncontiguous dates, multiple-range queries, and queries requiring aggregation of multiple results.

\begin{table}[H]
	\centering
	\caption{Examples of structured wearable data queries.}
	\label{tab:query-examples}
	\small
	\begin{tabular}{L{2.8cm} C{1.4cm} L{9.2cm}}
		\toprule
		\textbf{Query Type} &
		\textbf{Role ID} &
		\textbf{Example Question} \\
		\midrule
		
		Single date &
		5 &
		Total steps on March 16th, 2026. \\
		
		\addlinespace
		
		Continuous range &
		14 &
		How many times was atrial fibrillation detected between
		March 13th, 2026 and March 21st, 2026? \\
		
		\addlinespace
		
		Noncontiguous dates &
		2 &
		The total number of times I sat for long periods on
		March 8th, 9th, 10th, 12th, 17th, 19th, 27th, and
		29th, 2026. \\
		
		\addlinespace
		
		Multiple ranges &
		13 &
		Total calories for the 28 days:
		March 1--18 and March 21--30, 2026. \\
		
		\addlinespace
		
		Multiple-result aggregation &
		3 &
		On these 19 days: March 2, 5, 6, 7, 8, 9, 11, 12,
		13, 14, 15, 16, 17, 18, 21, 23, 25, 26, and 29,
		2026 --- how many times and how long did you exercise
		in total? \\
		
		\bottomrule
	\end{tabular}
\end{table}

These examples span daily activities and health events, as well as several temporal and output structures: single dates, continuous intervals, noncontiguous dates, multiple intervals, and aggregation of multiple results.

\subsection{Multi-Intent Question Examples}
\label{app:multi-intent-examples}

Table~\ref{tab:multi-intent-examples} presents six representative question types for the same synthetic user from the multi-intent recognition evaluation set. Each \texttt{role\_id} is associated with six predefined intent-type combinations:
\texttt{query, query},
\texttt{query, advice},
\texttt{query, analyse},
\texttt{query, query, query},
\texttt{query, query, advice}
and
\texttt{query, analyse, advice}.

These combinations evaluate whether the Manager Agent can identify multiple distinct user intents in a single complex health question and recover their types and multiplicities.

\begingroup
	\small
	\setlength{\tabcolsep}{3pt}
	\begin{longtable}{C{1.2cm} L{3.5cm} L{10.2cm}}
		\caption{Examples of multi-intent recognition questions.}
		\label{tab:multi-intent-examples} \\
		\toprule
		\textbf{Role ID} &
		\textbf{Intent Types} &
		\textbf{Example Question} \\
		\midrule
		\endfirsthead
		\caption[]{Examples of multi-intent recognition questions (continued).} \\
		\toprule
		\textbf{Role ID} &
		\textbf{Intent Types} &
		\textbf{Example Question} \\
		\midrule
		\endhead
		\midrule
		\multicolumn{3}{r}{\small Continued on next page} \\
		\endfoot
		\bottomrule
		\endlastfoot
		
		1 &
		\texttt{query, query} &
		Can you tell me what exercise sessions I did on 2026-03-09,
		including duration and calories burned for each, and also report
		my sleep duration and sleep score on 2026-03-08? \\
		
		\addlinespace
		
		1 &
		\texttt{query, advice} &
		Could you tell me which days in March I had detected premature
		heartbeats, and based on my stress-related headaches and caffeine
		intake, what advice can you give me to lower my premature heartbeat
		risk? \\
		
		\addlinespace
		
		1 &
		\texttt{query, analyse} &
		Can you give me my total steps and sleep score for 2026-03-01,
		and then analyze how my weekly exercise volume relates to my sleep
		quality across the whole month? \\
		
		\addlinespace
		
		1 &
		\texttt{query, query, query} &
		Can you tell me my running exercise duration and calories burned
		on 2026-03-01, my total sleep duration and sleep score on
		2026-03-08, and my total steps and sedentary periods on
		2026-03-15? \\
		
		\addlinespace
		
		1 &
		\texttt{query, query, advice} &
		Hey assistant, on 2026-03-15, how many total steps did I take?
		Also, what were my sleep onset time, total sleep duration, and
		sleep score that day? Based on my overall monthly activity and
		sleep patterns, what advice can you give me to improve my sleep
		quality? \\
		
		\addlinespace
		
		1 &
		\texttt{query, analyse, advice} &
		Can you tell me my total exercise time and average sleep score for
		March 1--7, analyze how the days with exercise compared to the
		no-exercise day affected my sleep, and advise whether I should
		schedule rest days differently to improve my sleep quality? \\
		
	\end{longtable}
\endgroup

Each example contains two or three distinct intent instances. Here, \texttt{query} denotes a data-retrieval request, \texttt{analyse} denotes a request to analyze available health data, and \texttt{advice} denotes a request for health-support recommendations.

The \texttt{Intent Types} column contains manually annotated reference intent types rather than the Task types generated during task decomposition. Accordingly, this evaluation measures the Manager Agent's multi-intent recognition, including coverage of the reference types and consistency of repeated-intent counts; subsequent task decomposition and dependency construction are outside its evaluation scope.

\subsection{Multidimensional Longitudinal Trend Examples}
\label{app:multi-dimensional-trend-examples}

Table~\ref{tab:multi-dimensional-trend-examples} presents the 10 representative questions used in the multidimensional longitudinal trend evaluation. They span one-, two-, and three-week intervals as well as a full month, and cover exercise, daily activity, sleep, and health-monitoring indicators.

\begingroup
	\footnotesize
	\setlength{\tabcolsep}{3pt}
	\begin{longtable}{C{1.2cm} C{2.3cm} L{11.4cm}}
	\caption{Examples of multidimensional longitudinal trend analysis questions.}
	\label{tab:multi-dimensional-trend-examples} \\
	\toprule
	\textbf{ID} &
	\textbf{Time Range} &
	\textbf{Example Question} \\
	\midrule
	\endfirsthead
	\caption[]{Examples of multidimensional longitudinal trend analysis questions (continued).} \\
	\toprule
	\textbf{ID} &
	\textbf{Time Range} &
	\textbf{Example Question} \\
	\midrule
	\endhead
	\midrule
	\multicolumn{3}{r}{\small Continued on next page} \\
	\endfoot
	\bottomrule
	\endlastfoot
	Q1 &
	1 week &
	What changes occurred in my daily step count over the past week?
	How did my daily exercise frequency change?
	How did my daily atrial fibrillation events change? \\
	
	\addlinespace
	
	Q2 &
	1 month &
	What changes occurred in my weekly total step count over the past month?
	How did my weekly exercise frequency change?
	How did my weekly premature heartbeat events change? \\
	
	\addlinespace
	
	Q3 &
	2 weeks &
	Over the past two weeks, how did my daily sleep apnea events change?
	How did my weekly total step count change?
	How did my daily sleep duration change? \\
	
	\addlinespace
	
	Q4 &
	1 month &
	Over the past month, how did my weekly exercise duration change?
	How did my weekly sleep quality score change?
	How did my weekly atrial fibrillation event frequency change? \\
	
	\addlinespace
	
	Q5 &
	3 weeks &
	Over the past three weeks, how did my daily premature heartbeat events change?
	How did my daily calories burned through exercise change?
	How did my daily sleep duration change? \\
	
	\addlinespace
	
	Q6 &
	1 month &
	Over the past month, what was the trend of my weekly step count?
	How did my weekly sedentary time change?
	Do these changes indicate a change in my activity level? \\
	
	\addlinespace
	
	Q7 &
	2 weeks &
	Over the past two weeks, how did my daily sleep duration and sleep quality change?
	At the same time, did my premature heartbeat events change? \\
	
	\addlinespace
	
	Q8 &
	1 month &
	Over the past month, how did my weekly exercise frequency, sleep quality,
	and cardiac health indicators change respectively?
	Considering these indicators together, do they suggest any changes in my overall
	health status? \\
	
	\addlinespace
	
	Q9 &
	3 weeks &
	Over the past three weeks, how did my daily atrial fibrillation events,
	premature heartbeat events, and sleep apnea events change?
	Do these abnormal events show an increasing trend? \\
	
	\addlinespace
	
	Q10 &
	1 month &
	Over the past month, what long-term trends can be observed in my exercise
	patterns, sleep patterns, and daily activity levels?
	Do these changes indicate any health risks that require attention? \\
	\end{longtable}
\endgroup

%% file: tables/quality_evaluation_rubric.tex
\begin{table}[htbp]
	\centering
	\caption{Rubric for automatic evaluation of overall response quality.}
	\label{tab:quality_evaluation_rubric}
	\small
	
	\setlength{\tabcolsep}{4pt}
	\renewcommand{\arraystretch}{1.15}
	
	\begin{tabular}{p{3.0cm} p{1.1cm} p{10.5cm}}
		\toprule
		\textbf{Dimension} &
		\textbf{Score} &
		\textbf{Evaluation Criterion} \\
		\midrule
		
		\multirow{5}{2.8cm}{\centering\textbf{Trustworthiness}}
		& 1 & No supporting health evidence is provided. \\
		& 2 & The supporting evidence is vague. \\
		& 3 & One supporting health indicator is provided. \\
		& 4 & Multiple supporting health indicators are provided. \\
		& 5 & Multiple supporting health indicators are provided with explicit reasoning linkage. \\
		\midrule
		
		\multirow{5}{2.8cm}{\centering\textbf{Transparency}}
		& 1 & Only the conclusion is provided without explanation. \\
		& 2 & A simple explanation is provided. \\
		& 3 & A partial evidence-to-conclusion rationale is presented. \\
		& 4 & A complete evidence-to-conclusion rationale is presented. \\
		& 5 & A complete evidence-to-conclusion rationale is presented with explicit evidence sources. \\
		\midrule
		
		\multirow{5}{2.8cm}{\centering\textbf{Actionability}}
		& 1 & No recommendation is provided. \\
		& 2 & A vague recommendation is provided. \\
		& 3 & A general recommendation is provided. \\
		& 4 & A specific recommendation is provided. \\
		& 5 & A personalized and executable recommendation is provided. \\
		\bottomrule
	\end{tabular}
	
\end{table}